\documentclass[12pt,showpacs,showkeys,amsmath,amssymb]{revtex4}
\usepackage{amsmath,amsfonts,amsthm,amscd,amssymb,latexsym}
\usepackage{bm}
\usepackage{dcolumn}
\usepackage{graphicx}
\usepackage{epstopdf}
\usepackage{color}
\usepackage{epsf}
\usepackage{epsfig}
\usepackage{graphicx, epic, eepic, color}

\newcommand{\beq}{\begin{equation}}
\newcommand{\eeq}{\end{equation}}

\def\@{\partial_}
\def\be{\begin{equation}}
\def\ee{\end{equation}}

\def\negenspace{\kern-1.1em}

\def\sqr#1#2{{\vcenter{\hrule height.#2pt\hbox{\vrule width.#2pt
height#1pt \kern#1pt \vrule width.#2pt}\hrule height.#2pt}}}

\begin{document}

\title{Dynamic Dark Energy from the Local Limit of Nonlocal Gravity}

\author{Javad \surname{Tabatabaei}$^{1}$}
\email{smjty25@gmail.com}
\author{Abdolali \surname{Banihashemi}$^{1}$}
\email{abdolali.banihashemi@gmail.com}
\author{Shant \surname{Baghram}$^{1,a}$}
\email{baghram@sharif.edu}
\author{Bahram \surname{Mashhoon}$^{2,3}$} 
\email{mashhoonb@missouri.edu}

\affiliation{
$^1$Department of Physics, Sharif University of Technology, Tehran 11155-9161, Iran\\
$^2$School of Astronomy,
Institute for Research in Fundamental Sciences (IPM),
Tehran 19395-5531, Iran\\
$^3$Department of Physics and Astronomy,
University of Missouri, Columbia,
Missouri 65211, USA \\
$^a$Corresponding Author
}

\date{\today \\  "~Essay written for the Gravity Research Foundation 2023 Awards for Essays on Gravitation}

\begin{abstract}
Nonlocal gravity (NLG), a classical extension of Einstein's theory of gravitation, has been studied mainly in linearized form. In particular, nonlinearities have thus far prevented the treatment of cosmological models in NLG. In this essay, we discuss the local limit of NLG and apply this limit to the expanding homogenous and isotropic universe. The theory only allows spatially flat cosmological models; furthermore, de Sitter spacetime is forbidden. The components of the model will have different dynamics with respect to cosmic time as compared to the standard $\Lambda$CDM model; specifically, instead of the cosmological constant, the modified flat model of cosmology involves a dynamic dark energy component in order to account for the accelerated phase of the expansion of the universe. 
\end{abstract}

\pacs{04.20.Cv}
\keywords{Gravitation}

\maketitle

The cosmological observations of cosmic microwave background (CMB) radiation~\cite{Planck:2018vyg} as well as large scale structure (LSS) surveys~\cite{BOSS:2016wmc} indicate that in terms of energy, we live in a dark universe with about $70\%$ dark energy, about $25\%$ dark matter, and only about $5\%$ visible matter. The  source of the two main dark sectors is still a mystery to us  since we only know about their gravitational effects.  These two components of the universe contribute to the standard model of cosmology only through gravitation. The standard model of cosmology known as $\Lambda$CDM is based on the cosmological principle and Einstein's theory of general relativity (GR). This model assumes that the cosmological constant $\Lambda$ is the cause of the accelerated universe~\cite{Weinberg:2013agg} and dark matter is due to a cold component (CDM). This theory is truly ignorant about its two important players and yet gives us acceptable results for describing the universe on large scales as a homogenous coarse-grained fluid. The standard model also gives a reasonable connection between the universe in its early times and the one we observe at low redshifts through the distribution of structures in the universe~\cite{Dodelson2020}.

The continuing failure of experiments to find the particles of dark matter naturally leads to the possibility that what appears as dark matter in astrophysics and cosmology is indeed an aspect of the gravitational interaction. While the endeavor to unmask the true identity of the dark sectors continues, there is a persistent notion that perhaps we do not know the gravitational interaction well enough even at the classical level.  Einstein's theory of gravitation has been well tested and quite successful on the scales of the solar system and isolated binary star systems~\cite{Will}; moreover, the detection of gravitational radiation is further evidence in support of GR in connection with binary systems~\cite{LIGOScientific:2016aoc, Akhshi:2021nsy}. However, ``dark" gravitational effects of unknown origin seem to appear on the scales of galaxies and beyond.  A starting point to understand general relativity more deeply would be to investigate the physical basis of the assumptions that underlie Einstein's elegant theory in order to align it with reality.

As a field theory of gravitational interaction, general relativity has been patterned after Maxwell's electrodynamics. Maxwell originally formulated the basic equations of electromagnetism in terms of the  electromagnetic fields $(\mathbf{E}, \mathbf{B})$ and their excitations in a material medium $(\mathbf{D}, \mathbf{H})$. The latter contains the response of the medium in terms of its polarizability and magnetizability, respectively; therefore, there are constitutive relations that connect these fields. In their simplest forms, we have $\mathbf{D} = \epsilon\, \mathbf{E}$, where $\epsilon$ is the electric permittivity of the medium and $\mathbf{B} = \mu \,\mathbf{H}$, where $\mu$ is the corresponding magnetic permeability. Even in Maxwell's time, observations pointed to the necessity of nonlocal and possibly nonlinear constitutive relations~\cite{Hop, Poi}. Indeed, history dependence  must be taken into account in the description of the properties of material media; for instance, magnetic materials generally exhibit hysteresis. In the process of averaging the response of the atomic medium, the memory of past events must be taken into account. The resulting nonlocal constitutive relations contain kernels that incorporate the atomic and molecular physics of the background medium~\cite{Jackson, L+L, HeOb}.  Similarly, it appears rather natural that in describing the large scale structure and evolution of the universe, averaging procedures may be necessary for cosmology.  This leads to the problem of averaging spacetime; in this connection,  see~\cite{VanDenHoogen:2017nyy} and the references cited therein. On the other hand, nonlocality could be an intrinsic feature of the universal gravitational interaction~\cite{BMB}.  

Physics is local within the frameworks of special and general theories of relativity~\cite{Einstein}. Here, locality means that the observer can make physical predictions on the basis of information contained in the local spacetime patch around itself; indeed, there is no need to keep track of the memory of past events. In extending Lorentz invariance to accelerated systems in Minkowski spacetime, Lorentz transformations are applied point by point along accelerated world lines. Acceleration, ignored locally, is taken into account only through the variation in the velocity of motion that appears in the pointwise Lorentz transformations. This locality principle is physically justified if the velocity of the accelerated system is essentially constant during an elementary act of measurement. For pointlike coincidences of classical point particles and rays of radiation, locality holds; however, according to the Huygens principle waves are intrinsically nonlocal. In connection with classical field measurements, Bohr and Rosenfeld have shown that spacetime averaging is necessary~\cite{BoRo}. Furthermore, acceleration can deflect the observer's path from a geometrically defined geodesic, and it may rotate the observer's frame, diverting the frame from how it should naturally evolve along the path via parallel propagation; hence, accelerated motion involves local invariant scales of length and time. The challenge here mainly arises when we ask how an accelerated observer can measure wave phenomena and also how quantum effects are manifested in accelerated frames.  It, therefore, appears that the past history of the accelerated motion should in general be taken into account through a nonlocal kernel that depends upon acceleration. The nonlocality of accelerated systems eventually leads to nonlocal special relativity theory~\cite{Mashhoon:2008vr}. 

The cornerstone of general relativity, namely, Einstein's principle of equivalence, is based on the local connection between gravitation and inertia, the same inertia that governs the dynamics of accelerated systems.  This circumstance suggests that gravity should be nonlocal as well. The most natural approach to a classical nonlocal generalization of GR would be to introduce history dependence into the theory through a nonlocal constitutive relation in close analogy with the nonlocal electrodynamics of media~\cite{Hehl:2008eu, Hehl:2009es}. To this end, we first need to express GR in a form that resembles electrodynamics. The key idea here is the introduction of an orthonormal tetrad frame field adapted to preferred observers in spacetime and the utilization of an extended geometric framework that employs both the Riemann curvature of spacetime and the Weitzenb\"ock torsion of the preferred frame field. That is, in the extended framework we have one spacetime metric tensor and two metric-compatible connections: the standard torsion-free symmetric Levi-Civita connection and the curvature-free Weitzenb\"ock connection defined via the preferred frame field. The Weitzenb\"ock connection~\cite{We} renders the spacetime manifold parallelizable.  In the framework of teleparallelism~\cite{Itin:2018dru}, two distant vectors are considered parallel if they have the same components with respect to their local preferred frame fields. It turns out that in this framework, Einstein's theory becomes the well-known teleparallel equivalent of general relativity (TEGR), which is the gauge theory of the four-parameter Abelian group of spacetime translations~\cite{Cho}. Therefore, TEGR, though nonlinear, is formally analogous to electrodynamics and then we can use our intuition about electrodynamics and modify TEGR. 

To introduce history dependence within the extended GR framework, let us consider an admissible system of spacetime coordinates $x^\mu$ in spacetime with metric
\begin{equation}\label{1}
ds^2 = g_{\mu \nu}(x)\,dx^\mu \, dx^\nu\,.
\end{equation}
In this gravitational field, free test particles and null rays follow the geodesics of the spacetime manifold. In our convention, Greek indices run from 0 to 3, Latin indices run from 1 to 3, and the signature of the metric is +2; moreover, we employ units such that $c = 1$. We assume the existence of a preferred set of observers with adapted tetrads $e^\mu{}_{\hat {\alpha}}(x)$ that are orthonormal, namely,
\begin{equation}\label{2}
g_{\mu \nu}(x) \, e^\mu{}_{\hat {\alpha}}(x)\, e^\nu{}_{\hat {\beta}}(x)= \eta_{\hat {\alpha} \hat {\beta}}\,.
\end{equation}
Here,  indices without hats are normal spacetime indices,  hatted indices refer to the tetrad axes in the local tangent space, and $\eta_{\alpha \beta} = {\rm diag}(-1,1,1,1)$ is the Minkowski metric tensor.

The curvature-free Weitzenb\"ock connection is defined in terms of the tetrad frame field as 
\begin{equation}\label{3}
\Gamma^\mu_{\alpha \beta}=e^\mu{}_{\hat{\rho}}~\partial_\alpha\,e_\beta{}^{\hat{\rho}}\,.
\end{equation}
It simply follows from this definition that $\nabla_\nu\,e_\mu{}^{\hat{\alpha}}=0$, where $\nabla$ denotes covariant differentiation via the Weitzenb\"ock connection. In this way, the preferred tetrad frames are parallel throughout spacetime and provide a natural scaffolding. The  \emph{torsion} tensor corresponding to the Weitzenb\"ock connection is the gravitational field strength within the framework of teleparallelism and is given by
\begin{equation}\label{4}
C_{\mu \nu}{}^{\alpha}=\Gamma^{\alpha}_{\mu \nu}-\Gamma^{\alpha}_{\nu \mu}=e^\alpha{}_{\hat{\beta}}\Big(\partial_{\mu}e_{\nu}{}^{\hat{\beta}}-\partial_{\nu}e_{\mu}{}^{\hat{\beta}}\Big)\,.
\end{equation}
A remark is in order here regarding the analogy between the gravitational field strength $C_{\mu \nu}{}^{\hat{\alpha}}$ and the electromagnetic field strength $F_{\mu \nu} = \partial_\mu A_\nu - \partial_\nu A_\mu$; that is, for each ${\hat{\alpha}}={\hat{0}}, {\hat{1}}, {\hat{2}}, {\hat{3}}$ in the torsion tensor
\begin{equation}\label{5}
C_{\mu \nu}{}^{\hat{\alpha}}=e_\rho{}^{\hat{\alpha}}C_{\mu \nu}{}^{\rho}= \partial_{\mu}e_{\nu}{}^{\hat{\alpha}}-\partial_{\nu}e_{\mu}{}^{\hat{\alpha}}\,,
\end{equation}
we have an analogue of the electromagnetic field tensor defined in terms of the vector potential $e_{\mu}{}^{\hat{\alpha}}$. The extended GR framework also contains the \emph{contorsion} tensor
\begin{equation}\label{6}
K_{\mu \nu}{}^{\alpha} =\, ^{0}\Gamma^{\alpha}_{\mu \nu} - \Gamma^{\alpha}_{\mu \nu}\,,
\end{equation}
where $^{0}\Gamma^{\alpha}_{\mu \nu}$ denotes the symmetric Levi-Civita connection.  It follows from the metric compatibility of the Weitzenb\"ock connection that the contorsion tensor is related to the torsion tensor, namely,
\begin{equation}\label{7}
K_{\mu \nu \rho} = \frac{1}{2}\, (C_{\mu \rho \nu}+C_{\nu \rho \mu}-C_{\mu \nu \rho})\,.
\end{equation}
Therefore, the Levi-Civita connection given by the Christoffel symbol is the sum of the Weitzenb\"ock connection and the contorsion tensor. The Einstein tensor  and the gravitational field equations, via the Levi-Civita connection,  can then be expressed in terms of the teleparallelism framework resulting in the teleparallel equivalent of GR, namely, TEGR~\cite{BMB}. Einstein's field equations expressed in terms of torsion thus become the TEGR field equations
\begin{equation}\label{8}
\frac{\partial}{\partial x^\nu}\,\mathfrak{H}^{\mu \nu}{}_{\hat{\alpha}}+\frac{\sqrt{-g}}{\kappa}\,\Lambda\,e^\mu{}_{\hat{\alpha}} =\sqrt{-g}\,(T_{\hat{\alpha}}{}^\mu + \mathbb{T}_{\hat{\alpha}}{}^\mu)\,,
\end{equation}
where $\Lambda$ is the cosmological constant, $\kappa := 8 \pi G$, and the auxiliary torsion field $\mathfrak{H}_{\mu \nu \rho}$ is defined via the auxiliary torsion tensor $\mathfrak{C}_{\alpha \beta \gamma}$ as
\begin{equation}\label{9}
\mathfrak{H}_{\mu \nu \rho}:= \frac{\sqrt{-g}}{\kappa}\,\mathfrak{C}_{\mu \nu \rho}\,, \qquad \mathfrak{C}_{\alpha \beta \gamma} :=C_\alpha\, g_{\beta \gamma} - C_\beta \,g_{\alpha \gamma}+K_{\gamma \alpha \beta}\,.
\end{equation}
Here,  $C_\mu :=C^{\alpha}{}_{\mu \alpha} = - C_{\mu}{}^{\alpha}{}_{\alpha}$ is the torsion vector, $T_{\mu \nu}$  is the symmetric energy-momentum tensor of matter, and $\mathbb{T}_{\mu \nu}$ is the traceless energy-momentum tensor of the gravitational field 
\begin{equation}\label{10}
\mathbb{T}_{\mu \nu} := (\sqrt{-g})^{-1}\, (C_{\mu \rho \sigma}\, \mathfrak{H}_{\nu}{}^{\rho \sigma}-\tfrac{1}{4}  g_{\mu \nu}\,C_{\rho \sigma \delta}\,\mathfrak{H}^{\rho \sigma \delta})\,.
\end{equation}

Inspired by the electrodynamics of nonlocal media, it is possible to  introduce history dependence and equip TEGR with the memory of the past. In the resulting nonlocal theory of gravitation, the gravitational field would be locally defined but would satisfy partial integro-differential field equations. That is, we expect that the modified gravitational field equations would contain a certain average of the gravitational field over past events. Such a term should involve a weight  or ``kernel" function that in the case of electrodynamics is determined by the nature of the atoms and molecules of the medium. The main difference with the electrodynamics of nonlocal media has to do with the absence of the background atomic medium. Therefore, the nonlocal kernel of the theory in the case of gravitation must ultimately  be  determined by observation. 

We can view the relationship between $\mathfrak{H}_{\mu \nu \rho}$ and the torsion tensor in Equation~\eqref{9} as the local constitutive relation of TEGR.  In the nonlocal electrodynamics of media, the field equations of electrodynamics are  not changed, only the constitutive relation is made nonlocal. In a similar fashion, in nonlocal gravity (NLG), only the constitutive relation of TEGR is modified. This means that we replace $\mathfrak{H}$ in Equations~\eqref{8} and~\eqref{10} by $\mathcal{H}$ given by
\begin{equation}\label{11}
\mathcal{H}_{\mu \nu \rho} = \frac{\sqrt{-g}}{\kappa}(\mathfrak{C}_{\mu \nu \rho}+ N_{\mu \nu \rho})\,,
\end{equation}
where $N_{\mu \nu \rho} = - N_{\nu \mu \rho}$ is a tensor that is nonlocally related to the torsion tensor. That is, the components of $N_{\mu \nu \rho}$ measured by the preferred observers of the theory with adapted tetrads $e^\mu{}_{\hat{\alpha}}$ are associated with the corresponding measured components of $X_{\mu \nu \rho}$ that is directly connected to the torsion tensor, namely,
\begin{equation}\label{12}
N_{\hat \mu \hat \nu \hat \rho}(x) = \int \mathcal{K}(x, x')\,X_{\hat \mu \hat \nu \hat \rho }(x') \sqrt{-g(x')}\, d^4x'\,,
\end{equation}
where $\mathcal{K}(x, x')$ is the basic causal kernel of NLG and~\cite{BMB, Puetzfeld:2019wwo, Mashhoon:2022ynk} 
\begin{equation}\label{13}
X_{\hat \mu \hat \nu \hat \rho}= \mathfrak{C}_{\hat \mu \hat \nu \hat \rho}+ \check{p}\,(\check{C}_{\hat \mu}\, \eta_{\hat \nu \hat \rho}-\check{C}_{\hat \nu}\, \eta_{\hat \mu \hat \rho})\,.
\end{equation}
Here,  $\check{C}^\mu$ is the torsion pseudovector defined via the Levi-Civita tensor $E_{\alpha \beta \gamma \delta}$ by
\begin{equation}\label{14}
\check{C}_\mu :=\frac{1}{3!} C^{\alpha \beta \gamma}\,E_{\alpha \beta \gamma \mu}\,
\end{equation}
and $\check{p}\ne 0$ is a constant dimensionless parameter. The resulting theory of nonlocal gravity (NLG) is rather intricate and no exact solution is known except for the trivial result that the spacetime is Minkowskian in the absence of gravity.  Furthermore, de Sitter spacetime is not an exact solution of NLG~\cite{Mashhoon:2022ynk}. On the other hand, linearized NLG has been extensively studied~\cite{BMB}; in fact, in the Newtonian regime of NLG, it has been possible to account for the rotation curves of nearby spiral galaxies as well as for the solar system data~\cite{Rahvar:2014yta, Chicone:2015coa, Roshan:2021ljs, Roshan:2022zov}.  Thus far, the nonlinearity of NLG has made it impossible to find exact solutions for the strong-field regimes such as those involving black holes or cosmological models. However, it is possible that features of NLG survive in the local limit of the theory, just as in electrodynamics the local permittivity $\epsilon(x)$ and permeability $\mu(x)$ functions can preserve features of nonlocal electrodynamics of media.
 
The local limit of Equation~\eqref{12} can be obtained by assuming that the kernel is proportional to the 4D Dirac delta function, namely,
\begin{equation}\label{15}
\mathcal{K}(x, x') := \frac{S(x)}{\sqrt{-g(x)}}\,\delta(x-x')\,,
\end{equation}
where $S(x)$ is a dimensionless scalar function that must be determined based on observational data~\cite{Tabatabaei:2022tbq}.
In this case, the nonlocal constitutive relation~\eqref{11} reduces to $N_{\mu \nu \rho}(x) = S(x) X_{\mu \nu \rho}$; that is, 
\begin{equation}\label{16}
N_{\mu \nu \rho}(x) = S(x)\,[\mathfrak{C}_{\mu \nu \rho}(x) + \check{p}\,(\check{C}_\mu\, g_{\nu \rho}-\check{C}_\nu\, g_{\mu \rho})]\,
\end{equation}
and the constitutive relation takes the form
\begin{equation}\label{17}
\mathcal{H}_{\mu \nu \rho} = \frac{\sqrt{-g}}{\kappa}[(1+S)\,\mathfrak{C}_{\mu \nu \rho}+ S\,\check{p}\,(\check{C}_\mu\, g_{\nu \rho}-\check{C}_\nu\, g_{\mu \rho})]\,.
\end{equation}
 Here, the susceptibility function $S(x)$ is a characteristic of the background spacetime just as $\epsilon(x)$ and $\mu(x)$ are features of the medium in electrodynamics. For $S(x) = 0$, we recover TEGR; otherwise, we have a generalization of GR that we expect may retain features of NLG through $S(x)$. Indeed, Equation~\eqref{17} implies that to have GR as a limit, we must impose the requirement that $1+S > 0$. In this local limit of nonlocal gravity, explicit deviations from the locality have vanished; however, nontrivial aspects of NLG may have survived through $S(x)$ that would be interesting to study. We, therefore, explore the cosmological implications of this local limit of NLG. 

The very first application of the local limit of NLG to the expanding universe has a dramatic result and implies spatial flatness. We begin with the FLRW metric for a dynamic (expanding or contracting) universe that is homogeneous and isotropic and assume that $S$ is only a function of \emph{cosmic time} $t$  with $dS/dt \ne 0$. That the susceptibility $S$ should be time dependent is consistent with the assumption of a dynamic universe model. The field equations of the local limit of NLG imply that the only model consistent with these assumptions is the spatially \emph{flat} model.  Therefore, we assume a metric of the form
\begin{equation}\label{18}
ds^2 = -dt^2 + a^2(t)\,\delta_{ij}\,dx^i\,dx^j\,,
\end{equation}
where $a(t)$ is the scale factor. The gravitational field equations imply that our modified flat model is governed by 
\begin{equation}\label{19}
3(1+S) \left( \frac{\dot{a}}{a}\right)^2 = \Lambda + 8\pi G \rho\,
\end{equation}
and
\begin{equation}\label{20}
2(1+S) \frac{\ddot{a}}{a} + (1+S) \left( \frac{\dot{a}}{a}\right)^2 = \Lambda - 8\pi G P - 2 \frac{dS}{dt} \frac{\dot{a}}{a}\,,
\end{equation}
where $\rho(t)$ and $P(t)$ are the energy density and pressure of the background cosmic medium, respectively. Moreover, we recall that $1+S > 0$; however, to ensure that this relation holds as $S$ varies with cosmic time, we further assume that $dS/dt > 0$.  Next, differentiating Equation~\eqref{19} with respect to cosmic time $t$ and using Equation~\eqref{20}, we find
\begin{equation}\label{21}
\frac{d\rho}{dt} = - 3(\rho + P) \frac{\dot{a}}{a} - \frac{3}{8 \pi G} \frac{dS}{dt}\left( \frac{\dot{a}}{a}\right)^2\,,
\end{equation}
which implies, for $\rho + P\ge 0$ and $dS/dt >0$, that $\rho$ monotonically decreases as the universe expands. 

Equations~\eqref{19} and~\eqref{20} govern the dynamics of the modified flat model of cosmology. In Equation~\eqref{19}, $1+S$, $\Lambda$, and $\rho$ are all positive; hence, the universe expands forever and as $t \to \infty$, $a(t)$ approaches infinity,  while $\rho$ and $P$ approach zero. It appears that the universe is then dominated by the cosmological constant $\Lambda$; however, the variation of $S$ with cosmic time leads to inconsistency.

Let us assume that our model is dominated by the cosmological constant as $t \to \infty$ and $\rho = 0$. Therefore, Equation~\eqref{19} implies
\begin{equation}\label{22}
\frac{\dot{a}}{a} = (\Lambda/3)^{1/2} (1+S)^{-1/2}\,.
\end{equation}
Taking the derivative of this relation with respect to time, we find
\begin{equation}\label{23}
\frac{\ddot{a}}{a} - \left( \frac{\dot{a}}{a}\right)^2 = - \frac{1}{2}(\Lambda/3)^{1/2} (1+S)^{-3/2}\,\frac{dS}{dt}\,.
\end{equation}
Let us multiply this relation by $2(1+S) > 0$ and subtract  the result  from Equation~\eqref{20} to get
\begin{equation}\label{24}
3(1+S) \left( \frac{\dot{a}}{a}\right)^2 = \Lambda - 2 \frac{dS}{dt} \frac{\dot{a}}{a} + (\Lambda/3)^{1/2} (1+S)^{-1/2}\,\frac{dS}{dt}\,.
\end{equation}
Plugging Equation~\eqref{19} in Equation~\eqref{24} results in
\begin{equation}\label{e4}
2\frac{\dot{a}}{a}\frac{dS}{dt} = (\Lambda/3)^{1/2} (1+S)^{-1/2}\,\frac{dS}{dt}\,, 
\end{equation}
which contradicts Equation~\eqref{22}, since $dS/dt$ does not vanish. Therefore, de Sitter spacetime is not a solution of our modified flat model. This circumstance is in conformity with the fact that de Sitter spacetime is not a solution of nonlocal gravity (NLG)~\cite{Mashhoon:2022ynk}. The modified flat cosmological model under consideration here will never asymptotically approach a de Sitter phase. 

To conclude our discussion of the local limit of nonlocal gravity (NLG), we have shown that the first approximation of the NLG, which is its local limit, has an important effect on the evolution of the densities of constituents of the universe with cosmic time.  All of the components of the universe evolve differently from their standard $\Lambda$CDM prediction. Accordingly, the modified flat model replaces the cosmological constant with a  dark energy component with an evolving equation of state. An important part of future studies in cosmology is related to the idea to see if the cause of the accelerated expansion of the universe is indeed due to a cosmological constant or not. The fundamental nonlocal feature of gravitational interaction predicts that we will find dynamic dark energy in upcoming decades instead of a cosmological constant~\cite{Annis:2022xgg,Ferraro:2022cmj}.


\end{document}